# Bias dependent inversion of tunneling magnetoresistance in Fe/GaAs/Fe tunnel junctions

J. Moser, M. Zenger, C. Gerl, D. Schuh, R. Meier, P. Chen, G. Bayreuther , W. Wegscheider, and D. Weiss

Institut für Experimentelle und Angewandte Physik, Universität Regensburg, D-93040 Regensburg, Germany

C.-H. Lai

Department of Materials Science and Engineering, National Tsing Hua University, 300 Hsinchu, Taiwan, Republic of China

R.-T. Huang

Institute of Material Engineering, National Taiwan Ocean University, 202 Keelung, Taiwan, Republic of China

M. Kosuth and H. Ebert

Department Chemie und Biochemie, Universität München, D-81377 München, Germany

We investigated spin dependent transport through Fe/GaAs/Fe tunnel junctions. The tunneling magnetoresistance effect (TMR) was probed for different types of Fe/GaAs interfaces. For interfaces cleaned by hydrogen plasma the TMR effect is increased and observable at room temperature. If an epitaxial Fe/GaAs(001) interface is involved, the tunnel junction exhibits a bias dependent inversion of the TMR effect. This is a first experimental signature for band structure effects at a Fe/GaAs interface and relevant for spin injection experiments.

The use of ferromagnetic metals with a high Curie temperature, e.g. iron, is a promising option for spin injection into semiconductors. By growing epitaxial contacts on GaAs[1] or AlGaAs[2] an injected spin polarization of up to 32% was observed so far.[3] In these experiments the limitations due to the conductivity mismatch between Fe and the semiconductor[4] was overcome by the Schottky barrier at the Fe/GaAs interface.[5,6] An alternative route to get information about spin polarized currents across the Fe/GaAs interface is to use the TMR across Fe/GaAs/Fe tunnel elements.[7] Due to the thin oxide layers at the interfaces between Fe and GaAs, caused by our preparation technique, a spin polarization of the tunneling current of only up to 9% was observed.[8] Here we describe the TMR effect for three different types of Fe/GaAs interfaces and show (i) that the TMR effect can be enhanced by a factor of 4 and is *observable at room temperature* (RT) by reducing the oxide layer using hydrogen plasma and (ii) that *the TMR effect changes sign as a function of bias voltage if one of the Fe/GaAs interfaces is fully epitaxial.*

In our experiments three different types of Fe/GaAs interfaces were used: epitaxial interfaces, labeled 'id', oxidized interfaces ('ox') and interfaces where the oxide was reduced by hydrogen plasma ('hy'). As described previously we started from GaAs (001) heterostructures, grown by molecular beam epitaxy (MBE), which contain the 8 nm thin GaAs tunneling barrier sandwiched between two $Al_{0.72}Ga_{0.28}As$ sacrificial layers.[7] With optical lithography and highly selective wet chemical etching the sacrificial layers were removed to open a window for the deposition of iron. These steps need to be done for both sides of the GaAs barrier. To obtain different switching fields one of the Fe layers is covered with a cobalt film. Before deposition of iron the GaAs surface is exposed to air in this case and a native oxide layer forms. The deposition of iron was carried out by magnetron sputtering at a base pressure of $p < 8 \cdot 10^{-10}$ mbar. Finally, the iron (17 nm) and iron (13 nm)/cobalt (50 nm) layers were covered with a 100 nm gold layer. A schematic cross section of a complete device is shown in

Fig. 1a. In previous experiments[7] both the bottom interface (bi) and the top interface (ti) of the GaAs barrier were covered with a thin oxide layer; the maximum TMR ratio was 1,7%,[8] corresponding, within Jullière's model,[9] to a spin polarization of 9%. Jullière's model is adequate here as the interfaces are disordered and symmetry related matching of the electron Bloch waves across GaAs (see, e.g., Ref.[10]) is not expected.

As the small TMR observed so far is obviously related to the oxidized interface we tried to remove the oxide. This constitutes the second type of interface explored ('hy'): We applied hydrogen plasma before iron deposition. The $H^+$-plasma reacts with the oxides and hydrocarbons on the GaAs surface and forms volatile compounds. Contaminations on a GaAs surface can thus be effectively removed.[11,12] We used an ion sputter gun under an angle of ~22° with an acceleration voltage of 1 kV for 30 min at RT. The hydrogen treatment does not attack the GaAs barrier. This is demonstrated in Fig.1c, displaying an AlGaAs/GaAs/Fe stack after $H^+$ cleaning. The GaAs barrier still has the nominal thickness of 8 nm.

The third interface type is the one with Fe grown on a clean GaAs ('id') (001) surface, applicable only for one interface ('bi' in Fig. 1a). For that a freshly grown GaAs heterostructure was transferred by an UHV-'shuttle' from the MBE chamber into the magnetron sputtering system without breaking the UHV. The Fe is epitaxially grown at RT in the UHV-sputter chamber. In Fig.1 atomically resolved cross sectional transmission electron micrographs of an oxide free 'perfect' Fe/GaAs interface (Fig. 1d) and a $H^+$-plasma treated interface (Fig. 1e) are compared. While for the latter the interface is distorted over 2-3 monolayers the 'ideal' interface looks sharp on a monolayer scale. The lattice planes of the epitaxially grown iron layer can be clearly seen.

The TMR of the different devices was measured for temperatures T between 4.2 K and RT using either a variable temperature insert of a $^4$He-cryostat with a superconducting coil or, at RT, a copper coil to generate an in-plane magnetic field B. We employed the Semiconductor Analyzer HP 4155A to probe the resistance drop across the barrier in four-point configuration.

Fig. 2a shows the TMR effect at T=4.2 K and at RT for a tunnel junction with both the top (ti) and bottom interface (bi) cleaned by $H^+$. At 4.2 K, $\Delta R/R$ is ~ 5.6% corresponding to a spin polarization of 16.5% within Jullière's model. At room temperature still a clear TMR effect of 1.55% is observable, corresponding to a spin polarization of 8.8%. If the oxide is reduced by $H^+$ the total resistance drops. This is ascribed to the reduction of the native oxide. However, the resulting spin polarization is doubled compared to a junction with the oxide layers still present. The degree of spin polarization for three different non-epitaxial Fe/GaAs interface combinations is shown in Fig. 2b. All traces show monotonically decreasing polarization for increasing T. If only one side of the junction is cleaned with $H^+$ the corresponding average polarization P(T) runs between the two other traces. The origin of the polarization's T-dependence will be discussed elsewhere. *In any case the TMR effect is positive,* i.e. the resistance for parallel magnetization configuration is lower than for antiparallel alignment. For these samples always a monotonic decrease of TMR with increasing bias is found. If one interface is epitaxial and the other one is either oxidized or cleaned with $H^+$ the most interesting behavior emerges: At small bias the TMR effect is negative and changes sign for sufficiently high positive or negative bias voltage. This bias dependence for id-ox and id-hy interface combinations is displayed in Fig. 3. The TMR changes sign at $\sim -90$ mV and $\sim 400-500$ mV. For $U<0$ electrons are transmitted from the ideal to the disordered interface, at $U>0$ from the disordered to the epitaxial interface. The insets show corresponding magnetoresis-tance traces for positive and negative TMR. While the TMR is larger for the device with hy-interface the qualitative behavior is very similar for the sample with ox-interface.

Two mechanisms for inversion of TMR were discussed previously: resonant tunneling via localized states[13] or different polarity of the spin polarization of the two contacts.[14] The former mechanism is unlikely as the effect is not observed for the other interface combinations having the same barrier. A perfect epitaxial interface between iron and GaAs is hence crucial to observe inversion of the TMR. In Jullière's model $\Delta R / R_{\uparrow\uparrow}$ is given by $2P_1P_2/(1-P_1P_2)$ where $P_{1,2}$ is the spin polarization and $\Delta R = R_{\uparrow\downarrow} - R_{\uparrow\uparrow}$ the resistance difference for parallel and antiparallel magnetization of the contacts. The polarization P is the difference of majority and minority spin DOS at the Fermi energy $E_F$ divided by the total DOS at $E_F$. For the ox-ox, hy-ox, or hy-hy interface combinations the TMR is always positive and hence the spin polarization at both interfaces is either positive or negative. The bias dependence switching of the TMR polarity for devices involving one epitaxial interface corresponds, within this model, to a bias dependent switching of the spin polarization at the epitaxial Fe/GaAs interface. The epitaxial interface, though, excludes a straightforward application of Jullière's model. The band structure of iron and GaAs and hence transmission dependent on the in plane wave vector $\vec{k}_{\parallel}$ needs to be taken into account.[15-19] A similar bias dependent inversion of the TMR has been observed in Fe/MgO/Fe junctions[20] and was reproduced by ab-initio calculations.[21] Depending on U, size and polarity of the TMR change and the contributions of majority and minority spins vary. This is concordant with our observation. Thus our experiments give a first hint that the Fe and GaAs band structure together with $\vec{k}_{\parallel}$ dependent transmission determines the degree of spin injection at an epitaxial Fe/GaAs interface.

The experiments suggest that the spin polarization of the tunneling electrons changes as a function of energy. To check this we calculated the DOS at a Fe/GaAs interface layer by layer self-consistently within the framework of the local spin density (LSD) formalism.[22] For the band structure calculation itself the tight-binding version of the Korringa-Kohn-Rostoker (KKR) method [23,24] has been used. Indeed, as shown in Fig. 3, a reversal of polarisation is found at $E = -394\,\text{meV}$ for the iron monolayer closest to the GaAs barrier. In contrast to Tiusan et al.,[20] however, the bias dependence of the TMR could not be assigned in a straightforward manner to the reversal of spin polarization below $E_F$. One reason is the neglect of $\vec{k}_{||}$-dependent transmission coefficients. If we nevertheless assign the region of negative TMR to a polarization dominated by minority spins of the interface iron the following conclusions can be drawn: At sufficiently small bias spin injection is dominated by minority spins. At larger negative or positive bias the injection of majority spins takes over. This interpretation is also consistent with spin injection experiments using spin light emitting diodes.[3] These experiments were carried out at a reverse bias of $\sim -2\,\text{V}$ across the Fe/GaAs Schottky barrier and injected spins were identified as iron majority spins.[25] Also for lateral Fe/GaAs/Fe devices majority spin injection was observed for $U < -0.2\text{V}$.[26]

In summary our experiments suggest that the band structure matching at an epitaxial Fe/GaAs interface causes bias dependent switching of the TMR effect in Fe/GaAs/Fe tunneling elements.

We acknowledge illuminating discussions with I. Mertig and financial support by BMBF (grant 01BM464 NanoQuit) and by DFG (SFB 689).

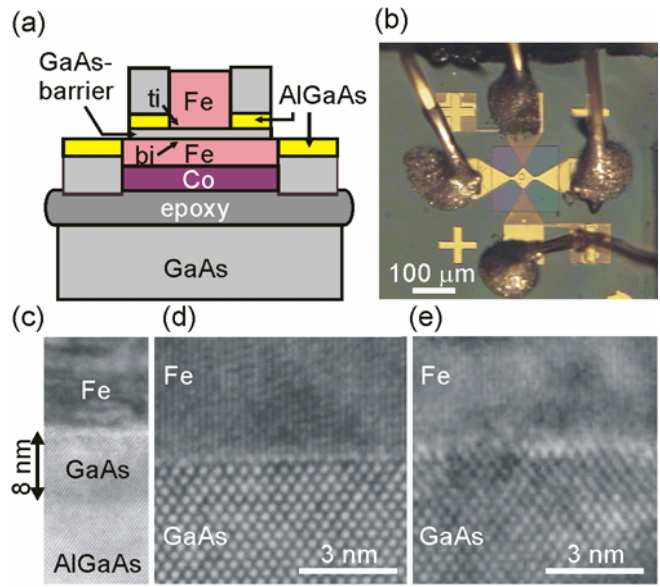

Fig. 1: a) Schematic cross section of a Fe/GaAs/Fe tunneling structure. b) Top view of a device with 4 wires attached. c) TEM micrograph showing the 8 nm thin GaAs barrier (cleaned by $H^+$) before preparation of the second Fe contact. d) High-resolution TEM micrograph of an ideal Fe/GaAs interface and e) of an interface after $H^+$-treatment.

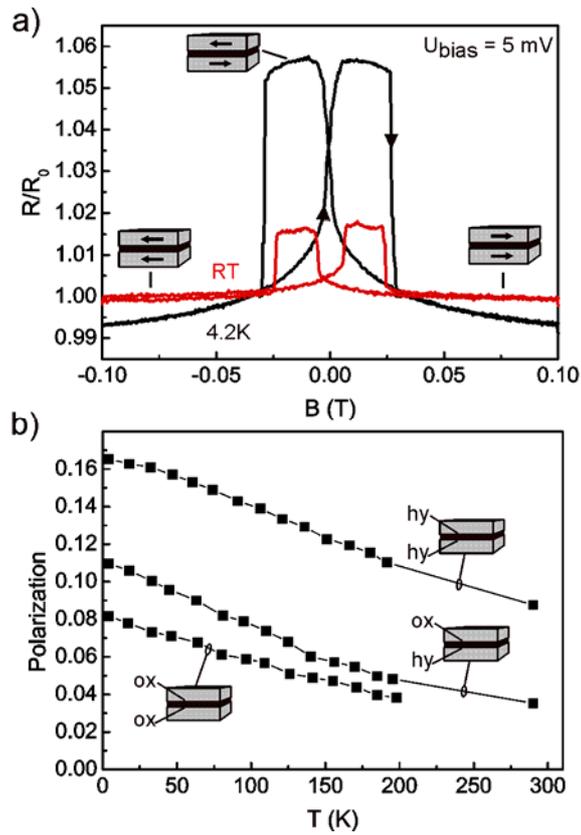

Fig. 2: a) Normalized TMR traces of a tunnel junction with 8 nm thick barrier at 4.2 K and at room temperature. Bias voltage: 5 mV. Both Fe/GaAs interfaces were cleaned with $H^+$ before deposition of iron. b) T-dependence of the spin polarization derived from Jullière's model for the interface combinations: hy-hy, ox-hy and ox-ox. Bias voltage: 5 mV.

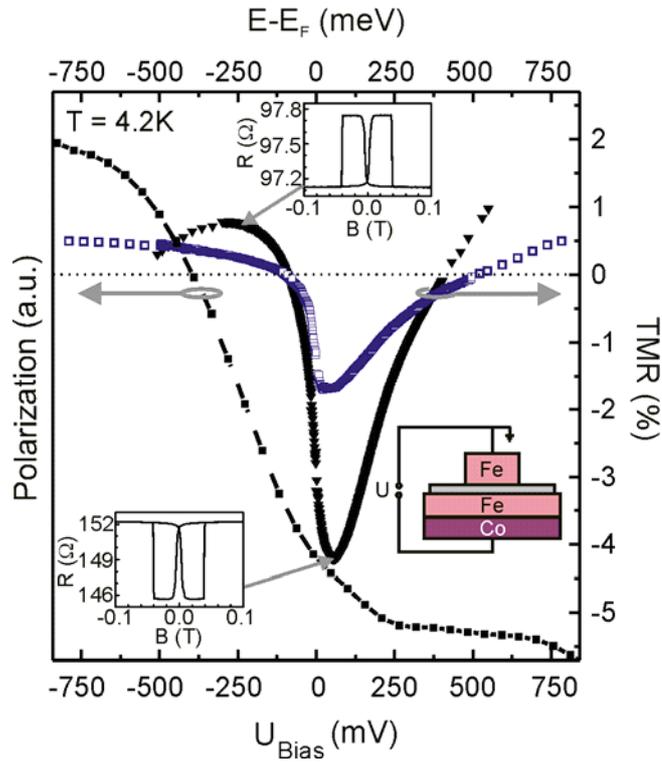

Fig. 3: (a) Bias dependence of the TMR for a Fe/GaAs(8nm)/Fe junctions with one 'ideal' epitaxial Fe/GaAs interface. The other interface was either untreated (open squares) or cleaned by a $H^+$-plasma (triangles). Two insets show examples for positive and negative TMR at U = -211 mV and U = +36 mV, respectively. The right inset displays the circuitry; the lower interface is the epitaxial one. Also shown is the calculated spin polarization for the Fe layer closest to the barrier. For energies above -394 meV minority spins dominate the interfacial density of states.